\documentclass[12pt,a4paper]{article}
\usepackage{cite}
\usepackage{comment}
\usepackage{graphicx}
\usepackage{tikz}
\usepackage{amsmath}
\usepackage{authblk}
\usepackage{hyperref}
\usepackage{mciteplus}
\usepackage{ifthen}

\newboolean{articletitles}
\setboolean{articletitles}{true} 

\begin{document}

\begin{titlepage}


\title{{\huge\bf A novel method to test particle ordering and final state alignment in helicity formalism}}
\author[1]{Mengzhen Wang}
\author[2]{Yi Jiang}
\author[2]{Yinrui Liu}
\author[2]{Wenbin Qian}
\author[2]{Xiaorui Lyu}
\author[1]{Liming Zhang\thanks{Corresponding author: liming.zhang@cern.ch}}

\affil[1]{Center for High Energy Physics, Tsinghua University, Beijing, China}
\affil[2]{University of Chinese Academy of Sciences, Beijing, China}

\maketitle

\vspace{\fill}
\begin{abstract}

In this article,  the importance is demonstrated of a proper choice of reference particles for decay angle definitions, when constructing partial-wave amplitude of multi-body decays using helicity formalism. 
This issue is often ignored in the standard use case of the helicity formalism.
A new technique is proposed to determine the correct particle ordering,  
and can also be used as a generalized method to calculate the rotation operators used for the final-state alignment between different decay chains. 
Numerical validations are also performed to support the arguments and new technique proposed in this article.  
\end{abstract}
\vspace{\fill}
\end{titlepage}

\def\Lb    {\Lambda_b^0}
\def\Km    {K^-}
\def\Lzst    {\Lambda^*}
\section{Introduction}
\label{sec:intro}
The partial-wave amplitude analysis is widely-used in experimental particle physics to understand the resonant structures in multi-body decays. By using the multi-dimensional phase-space variables, it gets more sensitivity to the properties of resonant states than a one-dimensional mass-spectrum analysis, and thus has become one of the most important techniques to explore  exotic hadrons and disentangle  contributions from  conventional contributions. Helicity formalism~\cite{Richman:1984gh} is one of the most popular techniques for constructing partial-wave amplitude. It sets a guideline to construct the angle-dependent amplitudes of two-body decays, which are further combined to form the amplitude of a decay chain (the cascade decay series made up of several two-body decays). The amplitudes of all the decay chains are combined to form the total amplitude of multi-body decays. When particles with non-zero spins are involved in the final state, a proper alignment of their spin axis should be made between different chains before the combination~\cite{Aaij:2015tga}, so that the final state of different decay chains are defined consistently. 

The helicity formalism has been used in the amplitude analysis of the $\Lambda_b^0\to \psi p K^-$ decay in the LHCb experiment, where the first observation of pentaquarks were made~\cite{Aaij:2015tga}. An alternative method for the amplitude construction of a three-body decay,  named as the Dalitz-Plot-Decomposition (DPD) approach, is proposed in Ref.~\cite{Mikhasenko:2019rjf}, and has been proved to be equivalent to the one used in the LHCb analysis~\cite{Mikhasenko:2019rjf}. However,  numerical comparisons show an unexpected dependency between the the consistency of these two formalisms and the choice of the reference particles when definiting the two-body decay angles, denoted the ``particle ordering issue''.  
Inspired by this observation, further investigations are made on the general rule for the decay amplitude construction using the helicity formalism.

 The rest part of the article will be organised as follows: Sec.~\ref{sec:puzzle} briefly introduces puzzles in the standard helisity formalism and the influence on the decay amplitude; 
 Sec.~\ref{sec:final-state-align} proposes a systematic method, based on the final-state alignment of different chains, to derive the correct helicity amplitude; Sec.~\ref{sec:numerical_final_state_alignment} shows an example use case on the $\Lambda_b^0 \to \psi p K^-$ amplitude analysis. 
 For a better clarification, the $\Lambda_b^0 \to \psi p K^-$ three-body decay is used as an example through the text, but the discussions and conclusions should validate for any other multi-body decays.

\section{Puzzles in the standard helicity formalism}
\label{sec:puzzle}
\begin{figure}[htbp]
  \begin{center}
    \includegraphics[width=0.8\linewidth]{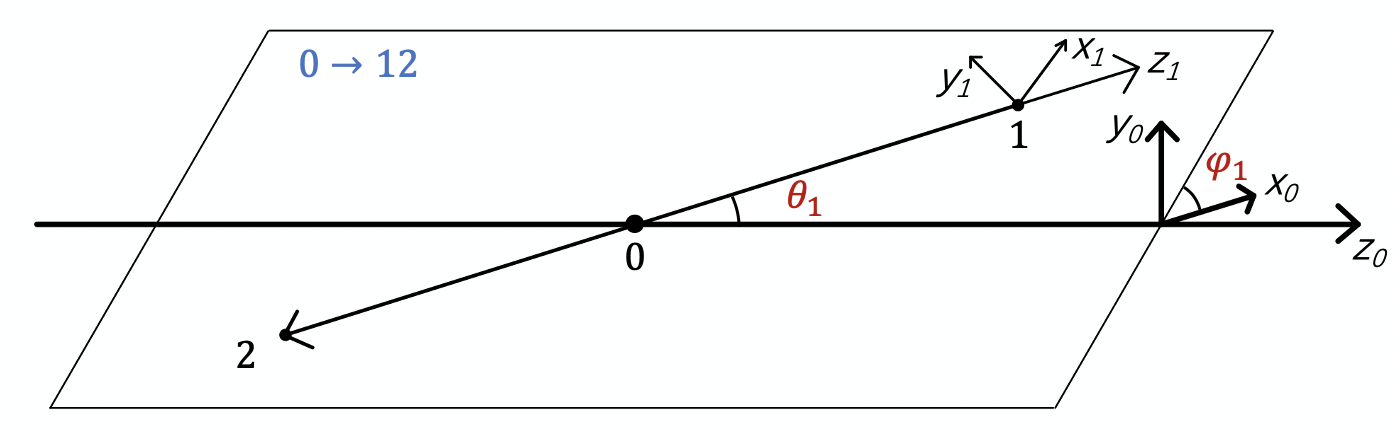}\put(-300,80){(a)}
    \vspace*{0.3cm}
    \includegraphics[width=0.8\linewidth]{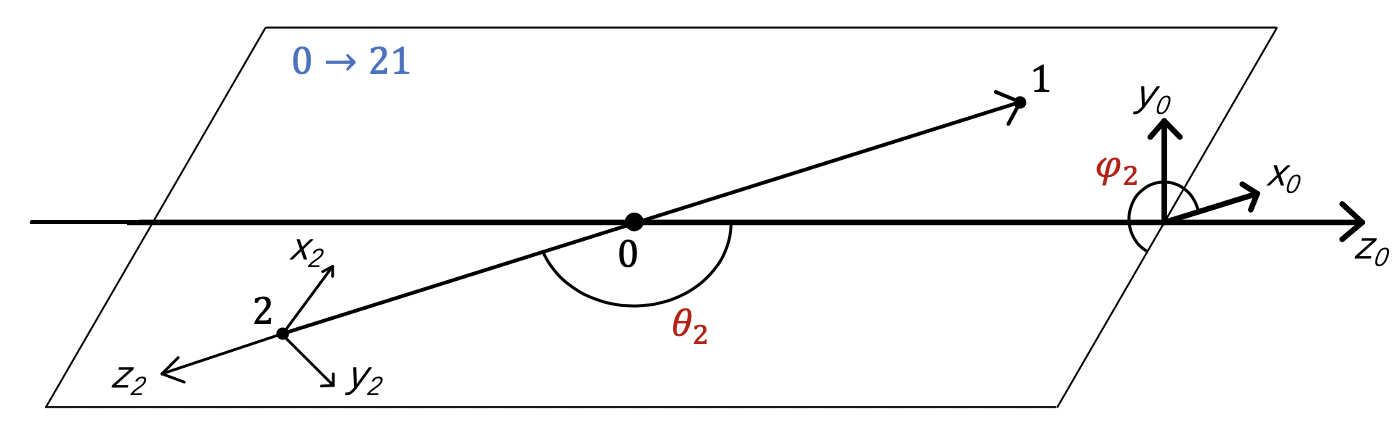}\put(-300,80){(b)}
    \vspace*{-0.5cm}
  \end{center}
  \caption{The helicity angles of the $0\to12$ decay (a) calculated based on the momentum of the particle 1, labelled as $(\theta_1, \phi_1)$ or (b) calculated based on the momentum of the particle 2, labelled as $(\theta_2, \phi_2)$.}
  \label{fig:helicity angle}
\end{figure}

The total amplitude of the $\Lambda_b^0 \to \psi p K^-$ decay is a sum of the amplitudes for the $\Lambda^*$ chain, namely  $\Lambda_b^0 \to \psi \Lambda^*, \Lambda^* \to p K^-$, and the $P_c$ chain, namely $\Lambda_b^0 \to P_c K^-, P_c \to \psi p$~\cite{Aaij:2015tga}. 
In the standard helicity formalism~\cite{Richman:1984gh}, the amplitude for any two-body decay listed above, denoted as $0\to12$, where 0, 1 and 2 indicate different particles, can be written as 
\begin{equation}
A_{0\to12}(m_0, \theta_1, \phi_1) =  H^{0\to12}_{\lambda_1, \lambda_2} D^{J_0}_{\lambda_0, \lambda_1-\lambda_2} (\phi_1, \theta_1, 0) R(m_0), 
\label{eq:helicity formalism}
\end{equation}
where $m_0$ denotes the invariant mass of the two-body system, $J_0$ is the  spin of particle 0. The angles  $\theta_1$ and $\phi_1$ stand for the polar and azimuthal angles of the momentum of particle 1 defined in the rest frame of particle 0, as displayed in Fig.~\ref{fig:helicity angle}, and also named as the helicity angles of the $0\to12$ two-body decay. 
Particle 1, whose momentum is used to define the helicity angles, is denoted hereafter the ``reference particle'' in this two-body decay. The label  $\lambda_i$ stands for the helicity of the particle $i$, which is defined as its spin projection onto the direction of the momentum.
The angle-dependent part of the amplitude is described using the Wigner D function $D^{J_0}_{\lambda_0, \lambda_1-\lambda_2}(\phi_1, \theta_1, 0)$~\cite{PDG2020}, corresponding to a rotation operator transfering the spin axis from the initial stage\footnote{Usually defined as the direction of the momentum of particle 0 in the rest frame of its originating particle, or in the lab frame is particle 0 is the starting particle of the entire decay chain}, to point to the momentum direction of particle 1.  The mass-dependency is denoted as a line-shape function $R(m_0)$, and a helicity coupling $H^{0\to12}_{\lambda_1, \lambda_2}$, which is a constant complex number, is used to describe the decay dynamics.

The total amplitude of the $\Lambda_b^0 \to \psi p K^-$ in  Ref.~\cite{Aaij:2015tga} is built following the standard procedure of the helicity formalism construction. The decay amplitude of the $\Lambda^*$ chain is the product of the amplitudes of the $\Lambda_b^0 \to \psi \Lambda^*$ and $\Lambda^* \to p K^-$ decays, and the decay amplitude of the $P_c$ chain is the product of the amplitudes of the $\Lambda_b^0 \to P_c K^-$ and $P_c \to \psi p$ decays. An alignment term, corresponding to a rotation within the decay plane, is added in the $P_c$ chain amplitude before the combination of the two decay chains. There are two potential defects remaining in this standard procedure. The first one, denoted the ``particle ordering issue'', is how to make a correct choice of reference particles in each two-body decay. And the second one, denoted the ``particle-two factor issue'', is about a properly description of the evolution on the spin axis, when the decay chain is connected using particles not taken as the reference of a two-body decay. These two issues are briefly mentioned in some other papers, such as Refs.~\cite{Mikhasenko:2019rjf, Marangotto:2019ucc}. In this article, more discussions are given to highlight the importance of these issues, in both analytic and numerical approaches. 

\subsection{Particle ordering issue}
\label{sec:single_node}
In the standard helicity formalism, there is in principle no preference on the choice of the reference particles. For the $0\to12$ decay, one should be able to take either particle 1 or particle 2 as the reference. If particle 2 is chosen, the angle-dependent part of the decay amplitude becomes $D^{J_0}_{\lambda_0, \lambda_2-\lambda_1}(\phi_2, \theta_2, 0)$. 
As shown in  Fig.~\ref{fig:helicity angle}, we have $\theta_1 + \theta_2 = \pi$, and $\phi_2 = \phi_2^\pm = \phi_1 \pm \pi$. The choice between $\phi_2^+$ and $\phi_2^-$ depends on how the range of the $\phi$ angles is defined. The most natural choice is to define both $\phi_1$ and $\phi_2$ in the same region, for example $[-\pi, \pi)$, and then  $\phi_2^+$ is taken when $\phi_1<0$, while $\phi_2^-$ is taken when $\phi_1>0$. Given the properties of the Wigner D functions, we have 
\begin{equation}
\label{eq:1_and_2_D}
D^{J_0}_{\lambda_0, \lambda_1-\lambda_2}(\phi_1, \theta_1, 0) = (-1)^{J_0+\lambda_0 \pm \lambda_0} D^{J_0}_{\lambda_0, \lambda_2-\lambda_1}(\phi_2, \theta_2, 0) , 
\end{equation}
and the difference of the angle-dependent amplitude when taking different reference particles is a factor of $f^{\pm} = (-1)^{J_0+\lambda_0 \pm \lambda_0}$, where $f^+$ is taken for decays with $\phi_1<0$, and $f^-$ is taken for decays with $\phi_1>0$. If particle 0 is a meson,  $f^{\pm}={(-1)}^{J_0}$ is just a global factor which can be absorbed by redefinition of helicity couplings. However, when particle 0 is a baryon, we have $f^{+} = -f^-$. The value of $f^{\pm}$ is different for events with $\phi_1 > 0$ or $\phi_1 < 0$. This minus sign has no effect on the module square of the amplitude of one decay chain. However, it  becomes non-negligible when multiple decay chains are considered, by directly influencing the behaviour of the interference terms, and it is impossible to eliminate this phase-space-dependent factor by introducing any global terms shared by all the events.

As an example, for the $\Lambda_b^0 \to \psi p K^-$ amplitude analysis, the reference particle of the $\Lambda^* \to p K^-$ decay is taken as the $K^-$ particle in Ref.~\cite{Aaij:2015tga}, and there is no particular reason of not taking the proton as the reference. Two possible settings of reference particles involved in the $\Lambda_b^0 \to \psi p K^-$ amplitude analysis are listed in Table.~\ref{tab:reference_particles}, where ``ordering 1'' corresponds to the choice of the LHCb analysis~\cite{Aaij:2015tga}, and ``ordering 2'' stands for the nominal choice used in the DPD paper which makes sure the decay planes for all the decay chains are identical in the ``aligned center-of-momentum frame'' ~\cite{Mikhasenko:2019rjf}. The only difference between these two orderings appears in decay angle definition of the $\Lambda^*\to p K^-$ decay, and they cannot be both correct, as discussed above, without introducing corrections which are handled differently when $\phi_{\Lambda^* \to p K^-}$ is larger or smaller than zero, where $\phi_{\Lambda^* \to p K^-}$ stands for the azimuth decay angle of this two-body decay. The technique for implementing this kind of correction is discussed later in Sec.~\ref{sec:another_use}. 

\begin{figure}[h]
\begin{center}
\includegraphics[width=0.45\textwidth]{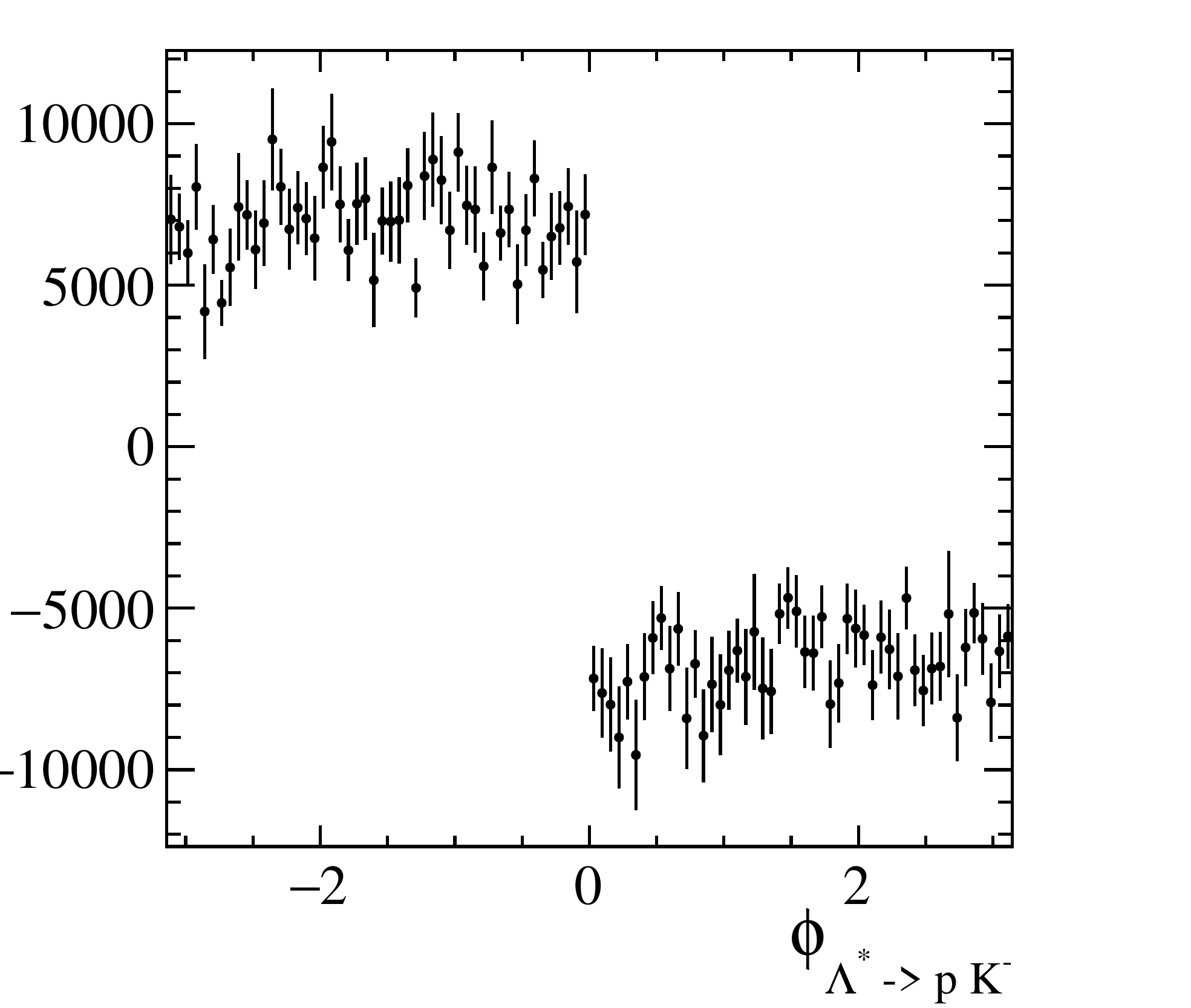}
\includegraphics[width=0.45\textwidth]{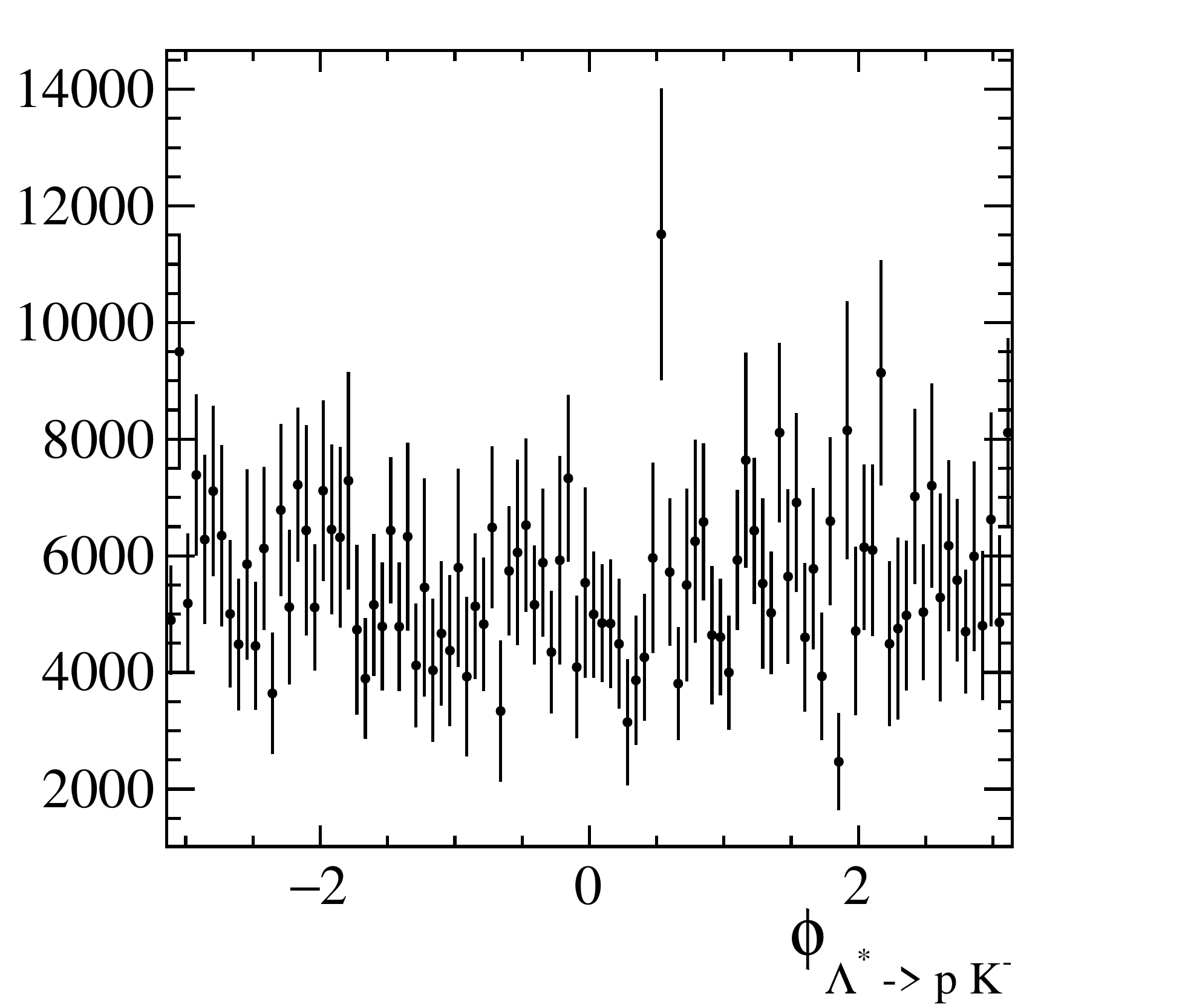}
\caption{The contribution from the interference term between the $\Lzst$ and $P_c$ chains as a function of $\phi_{\Lzst\to p \Km}$, obtained using simulated $\Lb \to \psi p \Km$ events. The left and right figures are based on particle ordering 1 and ordering 2, respectively. }
\label{fig:interference_contribution_phiK}
\end{center}
\end{figure}

The contributions of the estimated interference term between the amplitudes of the $P_c$ and $\Lambda^*$ chains, as a function of $\phi_{\Lambda^* \to p K^-}$,  calculated using the standard helicity formalism under ordering 1 and ordering 2, are shown in Fig.~\ref{fig:interference_contribution_phiK}. When ordering 1 is used, an unphysical discontinuity is seen at $\phi_{\Lambda^* \to p K^-} = 0$, indicating a potential problematic issue here, and the opposite behaviour of the interference term when $\phi_{\Lambda^* \to p K^-}>0$ and $\phi_{\Lambda^* \to p K^-}<0$ leads to a significant cancellation of the interference contribution when integrating the full  $\phi_{\Lambda^* \to p K^-}$ regions, which can explain the zero interference contribution between the $P_c$ and $\Lambda^*$ chains found in the ordering 1 configuration used in the LHCb analysis~\cite{Aaij:2015tga}.

\begin{figure}[htbp]
\begin{center}
\includegraphics[width=0.45\textwidth]{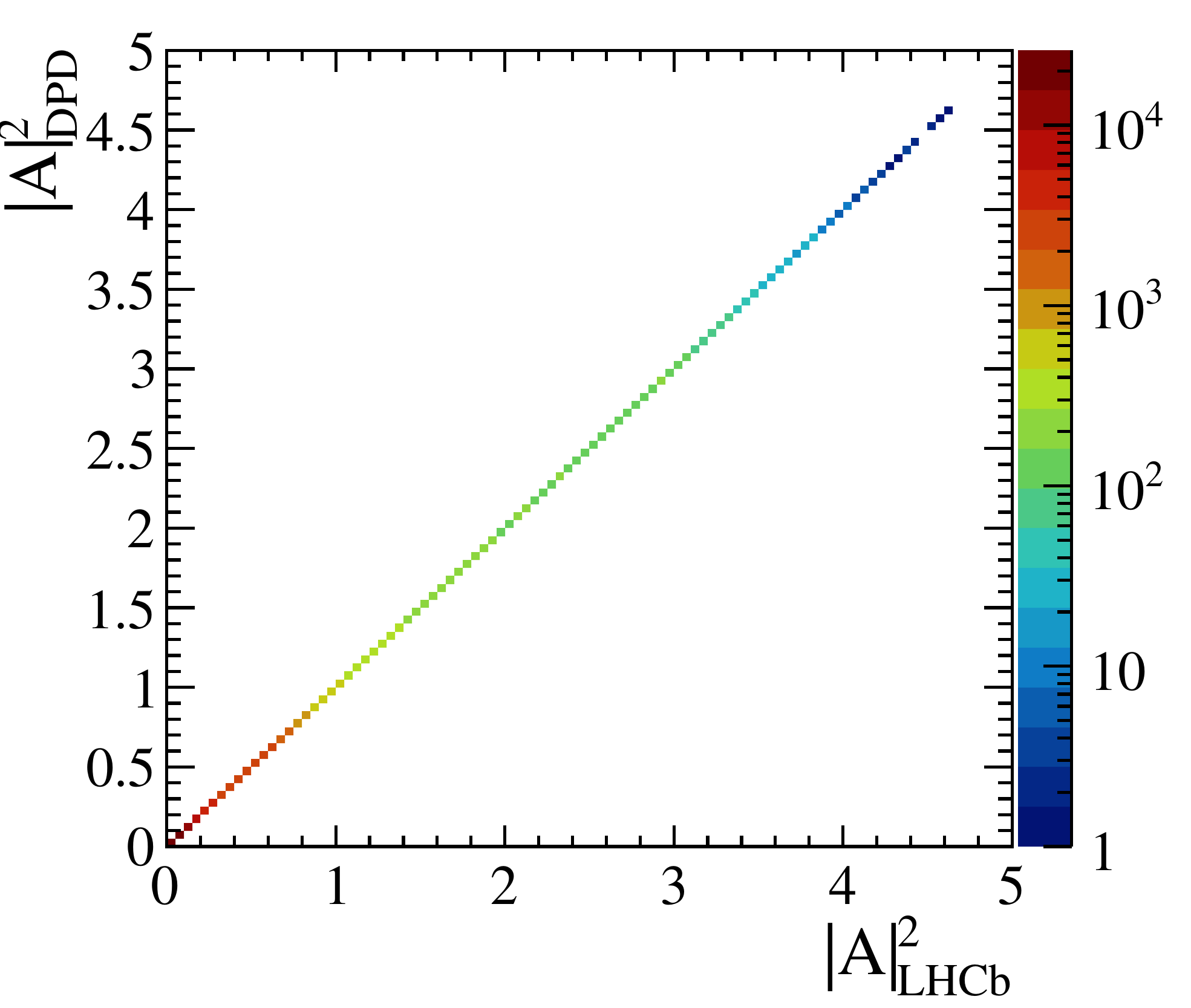}
\caption{Two dimensional distributions of the amplitude module square of the simulated $\Lb \to \psi p \Km$ decays based on ordering 2.  The x-axis stands for the amplitude module square calculated using the formula in Ref.~\cite{Aaij:2015tga}, while the y-axis is calculated using the DPD formula~\cite{Mikhasenko:2019rjf}. 
The mass-dependent terms and the helicity couplings are generated randomly and is shared in the two formulas.}
\label{fig:compare_2015_DPD_amplitudes}
\end{center}
\end{figure}

In the DPD formula~\cite{Mikhasenko:2019rjf}, the $\phi$-angle related terms are shared by all the decay chains, so it does not suffer from the potential non-global minus sign in the interference between different chains. A numerical comparison between the DPD formula and the standard helicity formalism helps to understand which ordering is correct.  
The decay amplitudes using both the two formulas are calculated for each simulated event. As shown in Fig.~\ref{fig:compare_2015_DPD_amplitudes}, these two formulas are equivalent when ordering 2 is used. It also indicates that ordering 1 needs further treatment.

\begin{table}[htbp]
\centering
\caption{The choice of the reference particles for each two-body decay involved in the $\Lambda_b^0 \to \psi p K^-$ amplitude analysis. Ordering 1 and ordering 2 correspond to the choices in Ref.~\cite{Aaij:2015tga} and Ref.~\cite{Mikhasenko:2019rjf}, respectively.}
\label{tab:reference_particles}
\begin{tabular}{ccc}
& & \\
\hline
 & Ordering 1 & Ordering 2 \\
 \hline
 $\Lambda_b^0 \to \psi \Lambda^*$ & $\Lambda^*$ & $\Lambda^*$ \\
 $\Lambda^* \to p K^-$ & $K^-$ & $p$ \\
 $\Lambda_b^0 \to P_c K^-$ & $P_c$ & $P_c$ \\
 $P_c \to \psi p$ & $\psi$ & $\psi$ \\
 \hline
\end{tabular}
\end{table}

\subsection{Particle-two factor issue}
\label{sec:particle-two-factor}
In the DPD paper~\cite{Mikhasenko:2019rjf}, the ``Jacob-Wick particle-2 phase convention'' is introduced. 
In this section, the particle-2 phase convention is explained in an alternative way, by introducing a ``particle-two factor'' to properly describe the transition of the spin axis. The effect on the particle parity determinations is also  demonstrated to support the necessity of including  this factor in the decay amplitude.

The rotation operator indicated in Eq.~\ref{eq:helicity formalism} results in a spin axis along the same direction as the momentum of particle 1. However, to consider the further decay of particle 2, the usually used initial spin axis is along the direction of the momentum of particle 2, which calls for an additional term between the amplitudes of $0\to1 2$ process and the further particle 2 decay. This additional term is named as the particle-two factor, and is generally ignored in the standard procedure to construct the helicity amplitude. 

To get the exact form of particle-two factor, let's define a Cartesian coordinate system $(x_1, y_1, z_1)$: as shown in Fig.~\ref{fig:helicity angle}, $\vec{z}_1$ and $\vec{z}_2$ are the directions of the momentum of particle 1 and 2, respectively, $\vec{y}_1$ is the normal vector of the $0\to12$ decay plane, and $\vec{x}_1 = \vec{y}_1 \times \vec{z}_1$. A rotation operator, corresponding to an angle of $\pi$, along any axis in the $x_1 - y_1$ plane should be able to transfer spin axis from $\vec{z}_1$ to $\vec{z}_2$.  Difference choices of the rotation axis result in different  $\vec{x}_2$ and $\vec{y}_2$ axis, leading to  different  definitions of the  helicity angles of the particle 2 decay, which are calculated using the directions of $\vec{z}_2$, $\vec{x}_2$ and $\vec{y}_2$ as inputs ~\cite{Aaij:2015tga}.  
A Wigner D function corresponds to this rotation operator should be taken as the particle-two factor. 

The most natural choice of the rotation axis is the $\vec{x}_1$ or $\vec{y}_1$ vector, both of which should be correct if the technique proposed later in Sec.~\ref{sec:final-state-align} is used. 
In Ref.~\cite{Aaij:2015tga}, the $\psi$ meson is not the reference particle in the $\Lambda_b^0 \to \Lambda^* \psi$ decay, and a rotation along the $\vec{x}_1$ axis is considered before calculating the $\psi\to\mu^+\mu^-$ decay angles. The corresponding particle-two factor,
\begin{equation}
\label{eq:particle-two-factor-old}
<J_2, -\lambda_2|R_x(\pi)|J_2, \lambda_2> = (-1)^{J_2},    
\end{equation}
is a constant parameter and can be absorbed into the definition of the helicity couplings. In this article, we suggest always taking the $\vec{y}_1$ axis as the rotation axis, which can avoid rolling over the decay planes to complicate the picture of helicity amplitude construction . The corresponding particle-two factor is 
\begin{equation}
\label{eq:particle-two-factor}
<J_2, -\lambda_2|R_y(\pi)|J_2, \lambda_2> = d_{-\lambda_2, \lambda_2}^{J_2} = (-1)^{J_2 - \lambda_2},    
\end{equation}
where $J_2$ and $\lambda_2$ are the spin and helicity of particle 2, respectively.

The particle-two factor is important to properly associate the helicity couplings and their $LS$ representations, which is discussed in Ref.~\cite{Mikhasenko:2019rjf}, and is also essential for the parity determination of the resonant states, as shown below.   
When particle 2 is a meson, this factor can be absorbed in the definition of the helicity coupling, namely  $H'^{0\to12}_{\lambda_1, \lambda_2} = (-1)^{J_2 - \lambda_2} H^{0\to12}_{\lambda_1, \lambda_2}$, and generates no visible effect in the amplitude analysis. If particle 2 is a baryon, the effect becomes non-negligible. The parity conservation requires that 
\begin{equation}
\label{eq:parity_conservation}
H^{0\to12}_{-\lambda_1, -\lambda_2} = P_0 P_1 P_2 (-1)^{J_1 + J_2 - J_0} H^{0\to12}_{\lambda_1, \lambda_2}, 
\end{equation}
where $P$ stands for the parity of the particles~\cite{Aaij:2015tga}. Under the parity transformation,  the particle-two factor varies from $(-1)^{J_2 - \lambda_2}$ to $(-1)^{J_2 + \lambda_2}$, generating an additional minus sign. One cannot absorb the particle-two factor of a baryon into the definition of the corresponding helicity coupling, whose behaviour under parity transformation would be otherwise modified, resulting in wrong determinations of particle parities.  

If the further decay of particle 2 is not considered in the amplitude analysis, we suggest also adding the particle-two factor after the $0\to12$ decay amplitude, to manage all the two-body decay amplitudes in a consistent way.  
Otherwise, it acts as an additional term for the final-state alignment between different chains.

\section{A technique for reference particle determination}
\label{sec:final-state-align}
As demonstrated in the previous section, the choice of the reference particle is not arbitrary,  when fermions are involved in the decay process.
It calls for a guideline to make the correct choice, or to add an additional term to the decay amplitude to cancel the non-global effect caused by switching the reference particles. Both of these two features are not well-described in the standard helicity formalism. 
A comparison between the standard helicity formalism and the DPD formula~\cite{Mikhasenko:2019rjf} is a possible approach, but it works only for three-body decays, and does not uncover the nature of this issue. In this section, we try to solve this puzzle by investigating the final-state alignment between different decay chains. 

In the $\Lambda_b^0 \to \psi p K^-$ amplitude analysis~\cite{Aaij:2015tga}, before the combination of the amplitude of the $\Lambda^*$ and $P_c$ chains, the spin state of the proton and muons should be properly aligned to be the same in different chains. 
The technique to correctly define the spin state of the decay products is raised in Ref.~\cite{Marangotto:2019ucc}, where the final spin states are obtained by carefully considering how they are associated to the initial spin state shared by all the decay chains.  
In this article, we focus on the operators connecting the initial and the final spin states, and propose an equation to define a correct final-state alignment in a mathematical way, which can be used to validate whether the alignment is handled properly when ordering 1 or 2 is taken. It can also be used as a generalized method to determine the correct alignment term in the decay amplitude, once the choice of the reference particles is fixed.

The target for the spin-axis alignment is to make sure that the combined $\Lambda^*$ chain and $P_c$ chain amplitudes are expressed indicating the same initial and final states. With a proper alignment, the rotation 
operators, which connect the initial and final spin states, should be identical in the $\Lambda^*$ and $P_c$ chains, and a mathematical description is 

\begin{equation}
\label{eq:align}
R_{{\rm Euler}, \Lambda^*}  = B_{ P_c} B^{-1}_{ \Lambda^*} R_{{\rm align}, P_c}  R_{{\rm Euler}, P_c}  
\end{equation} 
where the subscripts $\Lambda^*$ and $P_c$ are used to denote two decay chains, $R_{\rm Euler}$ stands for the cascade process of the Euler rotations associated to each two-body decays in the corresponding decay chain, with the Euler angles exactly taken from the decay angles in the amplitude. The symbol $R_{\rm align}$ stands for the alignment rotation for a consistent final state definition between $P_c$ and $\Lambda^*$ chains. Two boost operators, $B_{P_c}$ and $B^{-1}_{\Lambda^*}$, are involved to consider the difference between the reference frames under which the direction of the final spin axis is defined. The operator $B$ stands for the boost from the $\Lambda_b^0$ rest frame , to the rest frame of the final-state particles, through the intermediate resonances involved in each decay chain.

Once the particle ordering is determined, the decay angles of all the two-body decays can be calculated~\cite{Richman:1984gh,Aaij:2015tga}, based on which $R_{\rm Euler}$ can be obtained.
The boost operators can be calculated using the four-momentum information, and the alignment rotation becomes the only unknown part of Eq.~\ref{eq:align}.  By solving Eq.~\ref{eq:align} in either an analytic or a numerical way, one could determine the correct alignment rotations. If the alignment angle has been determined using other approaches~\cite{Aaij:2015tga}, Eq.~\ref{eq:align} can also be used to validate whether the alignment is properly performed.

As all the angles in Eq.~\ref{eq:align} are directly taken from the decay amplitude, it should be sensitive to the switch of the reference particles, and should generate visible effect once a proper representation is 
assigned on the rotation operators. 
For the $\Lambda_b^0 \to \psi p K^-, \psi \to \mu^+ \mu^-$ amplitude analysis, the final-state particles related to the puzzles in Sec.~\ref{sec:puzzle} are all spin-half states. 
So, in this article, the two-dimensional representation 
of the SU(2) group is used to describe the rotation operators. As the Wigner D function is the $jm$ representation of the SU(2) group, this should be a good choice to visualize the properties in Eq.~\ref{eq:1_and_2_D}.   The rotations along the z-axis, y-axis and an arbitary axis labelled as $\vec{a}$ are expressed using :
\begin{equation}
R_z(\alpha) =
\left(
\begin{array}{cc}
e^{-i\alpha/2} & 0 \\ 0 & e^{i\alpha/2} \\
\end{array}
\right), 
\end{equation}

\begin{equation}
R_y(\alpha) =
\left(
\begin{array}{cc}
\cos(\alpha/2) & -\sin(\alpha/2) \\ \sin(\alpha/2) & \cos(\alpha/2) \\
\end{array}
\right),
\end{equation}

and 
\begin{equation} 
R(\alpha, \vec{a}) = R_z(\phi_a)R_y(\theta_a)R_z(\alpha)R_y(-\theta_a)R_z(-\phi_a), 
\end{equation} 
respectively, where $\alpha$ stands for the rotation angle, with $\theta_a$ and $\phi_a$ denoting the polar and azimuthal angles of vector $\vec{a}$. 
The corresponding representation of the boost operators, inspired by the $(\frac{1}{2}, 0)$ representation of the Lorentz group~\cite{Srednicki:2004hg}, with a rapidity of $\gamma$, along the  z-axis and along any axis labelled as $\vec{b}$ are 
\begin{equation}
B_z(\gamma) =
\left(
\begin{array}{cc}
e^{-\gamma/2} & 0 \\ 0 & e^{\gamma/2} \\
\end{array}
\right), 
\end{equation}
and 
\begin{equation}
B(\gamma, \vec{b}) = R_z(\phi_b)R_y(\theta_b)B_z(\gamma)R_y(-\theta_b)R_z(-\phi_b), 
\end{equation} 
respectively, where $\theta_b$ and $\phi_b$ are the polar and azimuthal angles of vector $\vec{b}$.

\section{Validate different orderings for $\Lambda_b^0 \to \psi p K^-$ amplitude analysis}

\label{sec:numerical_final_state_alignment}

An example is shown in this section for the validation of alignment for the proton spin axis between the $P_c$ and $\Lambda^*$ chains. Following the methodology of calculating both the decay angles and the alignment angles in Ref.~\cite{Aaij:2015tga}, the two choices of reference particles listed in Table.~\ref{tab:reference_particles} are validated by checking if Eq.~\ref{eq:align} is satisfied. 
 
\subsection{Rotations when using ordering 1}
If ordering 1 is used,  the corresponding proton-related Euler rotation for the $\Lzst$ chain is 
\begin{equation}
\label{eq:ordering1_Lzst}
    R_{\rm Euler, \Lzst} = 
R(\pi, \vec{p}_{\Lambda^*}^{\Lb} \times \vec{p}_{K}^{\Lambda^*})
R(\theta_{\Lambda^*}, \vec{p}_{\Lambda^*}^{\Lb} \times \vec{p}_{K}^{\Lambda^*})  
    R(\phi_K, \vec{p}_{\Lambda^*}^{\Lb}) 
    R(\theta_{\Lb}, \vec{p}_{\Lambda^*}^{\Lb} 
    \times \vec{p}_{\Lb}^{lab}) R(\phi_{\Lambda^*}, \vec{p}_{\Lb}^{lab}) , 
\end{equation}
where $\theta_{\Lb}$ and $\phi_{\Lambda^*}$ are the helicity angles of the $\Lb \to \Lzst \psi$ decay, $\theta_{\Lambda^*}$ and $\phi_K$ are the helicity angles of the $\Lzst \to p \Km$ decay~\cite{Aaij:2015tga}, and the symbol $\vec{p}_a^b$ is used to denote the direction of the momentum of particle $a$ in the $b$ rest frame. The rotation operator 
$R(\theta_{\Lb}, \vec{p}_{\Lambda^*}^{\Lb} \times \vec{p}_{\Lb}^{lab}) R(\phi_{\Lambda^*}, \vec{p}_{\Lb}^{lab})$ corresponds to the helicity amplitude of the $\Lambda_b^0 \to \Lambda^* \psi$ decay, and it transfers the spin axis from the direction of the $\Lb$ momentum in the lab frame to that of the $\Lambda^*$ momentum in the $\Lambda_b^0$ rest frame. The operator  $R(\pi, \vec{p}_{K}^{\Lambda^*})  R(\theta_{\Lambda^*}, \vec{p}_{\Lambda^*}^{\Lb} \times \vec{p}_{K}^{\Lambda^*})$ corresponds to the helicity amplitude of the $\Lambda^* \to p K^-$ decay, and the direction spin axis is transfered along the direction of the $K^-$ momentum in the $\Lambda^*$ rest frame. An operator $R(\pi, \vec{p}_{\Lambda^*}^{\Lb} \times \vec{p}_{K}^{\Lambda^*})$ is added, corresponds to the particle-two factor described in Sec.~\ref{sec:particle-two-factor}, as proton is not the reference particle for the $\Lambda^* \to p K^-$ decay angle definition.  
Similarly, for the $P_c$ chain, the Euler rotation is 
\begin{equation}
\label{eq:ordering1_Pc}
R_{\rm Euler, P_c} =  
R(\pi, \vec{p}_{\psi}^{P_c} \times \vec{p}_{K}^{P_c})
R(\theta_{P_c}, \vec{p}_{\psi}^{P_c} \times \vec{p}_{K}^{P_c}) \\
R(\phi_\psi^{P_c}, \vec{p}_{P_c}^{\Lb}) R(\theta_{\Lb}^{P_c}, \vec{p}_{P_c}^{\Lb} \times \vec{p}_{\Lb}^{lab}) R(\phi_{P_c}, \vec{p}_{\Lb}^{lab}),
\end{equation}
where \mbox{$R(\theta_{\Lb}^{P_c}, \vec{p}_{P_c}^{\Lb} \times \vec{p}_{\Lb}^{lab}) R(\phi_{P_c}, \vec{p}_{\Lb}^{lab})$} and \mbox{ $R(\theta_{P_c}, \vec{p}_{\psi}^{P_c} \times \vec{p}_{K}^{P_c}) \\
R(\phi_\psi^{P_c}, \vec{p}_{P_c}^{\Lb})$ } correspond to the helicity amplitude of the $\Lambda_b^0 \to P_c K^-$ and $P_c \to \psi p$ decays~\cite{Aaij:2015tga}, respectively, and the spin axis is transfered from the same initial stage as that in the $\Lambda^*$ chain, along the direction of the $\psi$ momentum in the $P_c$ rest frame. The operator $R(\pi, \vec{p}_{\psi}^{P_c} \times \vec{p}_{K}^{P_c})$ corresponds to the particle-two factor. 
The alignment rotation is expressed in  
\begin{equation}
\label{eq:ordering1_align}
    R_{\rm align, P_c} =
    R(\pi, \vec{p}_{K}^{\Lambda^*})
    R(\theta_p, \vec{p}_{\psi}^{P_c} \times \vec{p}_{K}^{P_c}), 
\end{equation}
corresponding to the alignment term between the $\Lambda^*$  and $P_c$ chains used in Ref.~\cite{Aaij:2015tga}, where the alignment angle  $\theta_p$  is defined as the angle between the $K^-$ and $\psi$ particles in the proton rest frame and the value is restricted in $\theta_p\in(0,\pi)$ . When ordering 1 is used,  the normal directions of the decay planes are opposite in the $\Lzst$ and $P_c$ chains, so an additional rotation $R(\pi, \vec{p}_K^{\Lambda^*})$, with an angle of $\pi$ along the spin axis, is introduced to eliminate this difference. 

\subsection{Rotations when using ordering 2}
If ordering 2 is used, the Euler rotation for the $\Lzst$ chain becomes 
\begin{equation}
    \label{eq:ordering2_Lzst}
    R_{\rm Euler, \Lzst} = R(\theta_{\Lambda^*}, \vec{p}_{\Lambda^*}^{\Lb} \times \vec{p}_{p}^{\Lambda^*}) R(\phi_p, \vec{p}_{\Lambda^*}^{\Lb}) R(\theta_{\Lb}, \vec{p}_{\Lambda^*}^{\Lb} \times \vec{p}_{\Lb}^{lab})R(\phi_{\Lambda^*}, \vec{p}_{\Lb}^{lab}).
\end{equation}
The Euler rotation for the $P_c$ chain becomes 
\begin{equation}
    \label{eq:ordering2_Pc}
    R_{\rm Euler, P_c} = R(\pi, \vec{p}_{\psi}^{P_c} \times \vec{p}_{K}^{P_c}) R(\theta_{P_c}, \vec{p}_{\psi}^{P_c} \times \vec{p}_{K}^{P_c})   
R(\phi_\psi^{P_c}, \vec{p}_{P_c}^{\Lb}) R(\theta_{\Lb}^{P_c}, \vec{p}_{P_c}^{\Lb} \times \vec{p}_{\Lb}^{lab}) R(\phi_{P_c}, \vec{p}_{\Lb}^{lab}), 
\end{equation}
and the alignment rotation  $R_{\rm align, P_c}$ is 
\begin{equation}
\label{eq:ordering2_align}
    R_{\rm align, P_c} =
    R(\theta_p, \vec{p}_{\psi}^{P_c} \times \vec{p}_{K}^{P_c}). 
\end{equation}
The definition of all the angles are almost the same as that for ordering 1, except that $\theta_{\Lambda^*}$ and $\phi_p$, which are the $\Lzst \to p \Km$ decay angles defined using the proton, rather than the $K^-$ particle as the reference. 
The rotation operator $R(\pi, \vec{p}_{\psi}^{P_c} \times \vec{p}_{K}^{P_c})$ corresponds to the particle-two factor mentioned in Sec.~\ref{sec:particle-two-factor}. 

\subsection{Boost operators}
The boost operators are defined in the same way for both ordering 1 and ordering 2. The operator 
$B_{\rm \Lzst}$ first boosts the $\Lb$ rest frame to the $\Lzst$ rest frame, and then to the proton rest frame, namely
\begin{equation}
    B_{\rm \Lzst} = B(-y_{p}^{\Lzst}, \vec{p}_{p}^{\Lzst})  B(-y_{\Lzst}^{\Lb}, \vec{p}_{\Lzst}^{\Lb}), 
\end{equation}
where the symbol $y_{a}^{b}$ stands for the rapidity of particle a, 
\begin{equation}
    y = \frac{1}{2} \ln (\frac{E_a+p_a}{E_a-p_a}), 
\end{equation}
defined in the rest frame of particle b. 
Similarly $B_{\rm P_c}$ first boost the $\Lb$ rest frame to the $P_c$ rest frame, and then to the proton rest frame, namely 
\begin{equation}
    B_{\rm P_c} = B(-y_{p}^{P_c}, \vec{p}_{p}^{P_c})  B(-y_{P_c}^{\Lb}, \vec{p}_{P_c}^{\Lb}), 
\end{equation}

\subsection{Validate the alignment equation}
\begin{figure}[h]
\begin{center}
\includegraphics[width=0.45\textwidth]{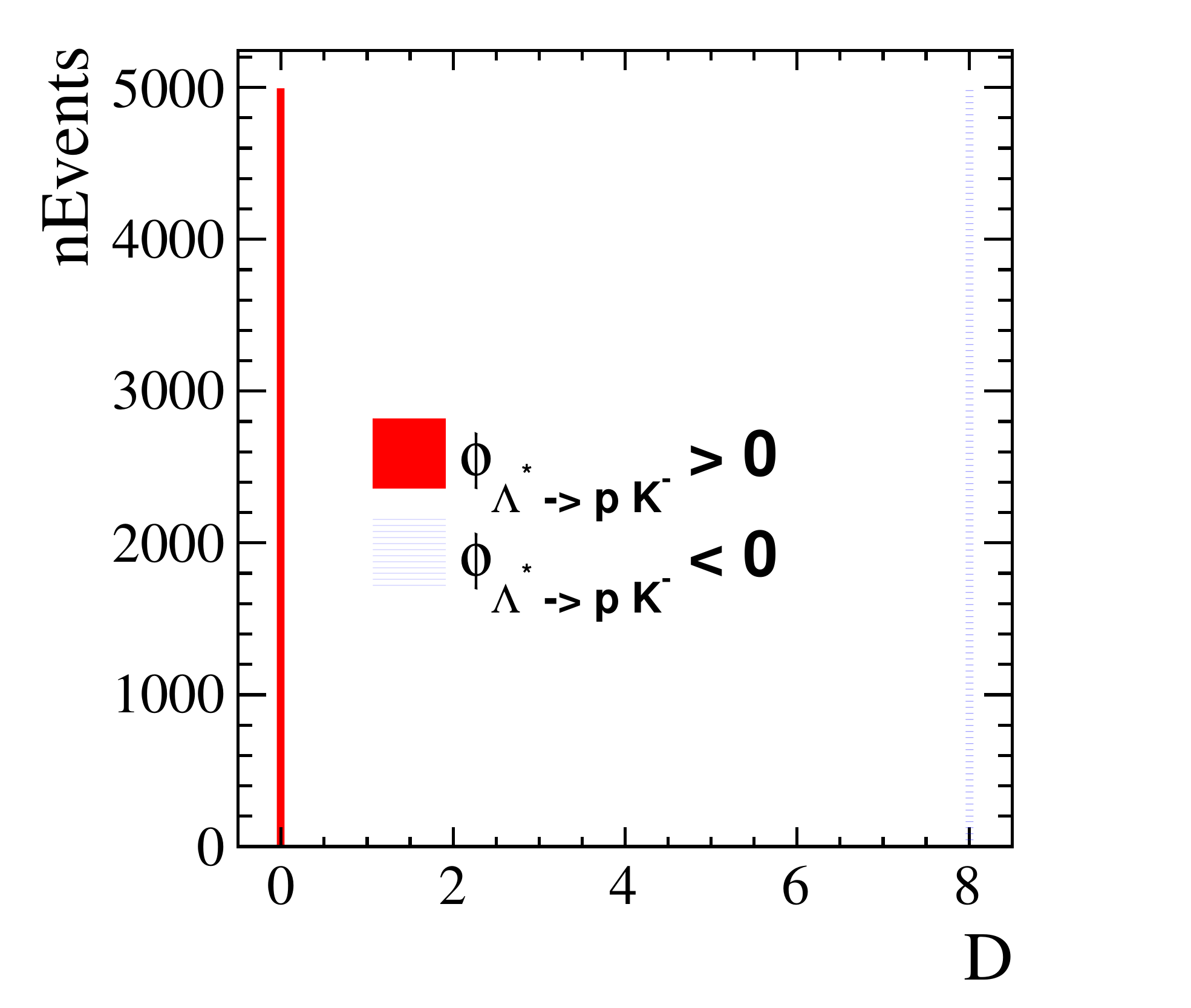}
\includegraphics[width=0.45\textwidth]{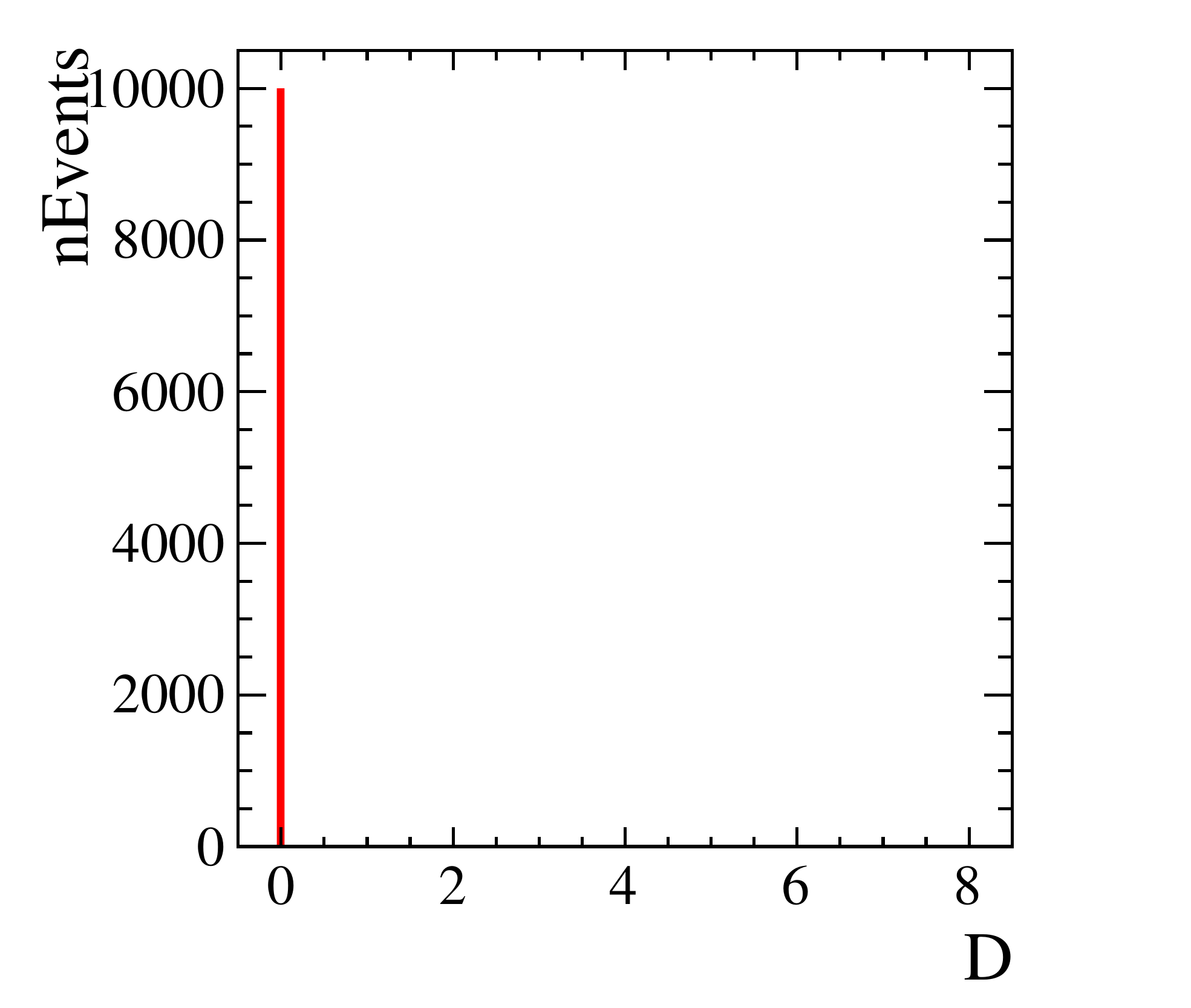}
\caption{The distribution of $D$ obtained with the helicity and alignment angles obtained using the method proposed in Ref.~\cite{Aaij:2015tga}. The figure in the left is obtained based on ordering 1, where half of the candidates have a perfect alignment between the two chains, while the other half $D=8$, corresponding to a minus sign difference between the rotation matrices of the two chains. The figure in the right is obtained ordering 2, where all the rotation matrices of the two chains are perfectly aligned for all the generated events.}
\label{fig:alignment}
\end{center}
\end{figure}

For a better visibility of validation of Eq.~\ref{eq:align}, the distance between the matrices in its left and right sides is defined as 
\begin{equation}
    D = \sum_{i,j} | L_{i,j}-R_{i,j}|^2 , 
\end{equation}
where $L$ stands for the left side, namely $R_{\rm Euler, \Lzst}$, while $R$ for the right side, namely $B_{P_c} B_{\Lzst}^{-1} R_{\rm align, P_c} R_{\rm Euler, P_c}$, and the subscripts $i$ and $j$ are the row and column indexes for $L$ or $R$ matrices, respectively. A distance $D$ of zero indicates that $L$ equals to $R$. As both $L$ and $R$ are $2\times2$ unitary matrices, when $L=-R$ the distance becomes $D=8$. 
Figure~\ref{fig:alignment} shows the distribution of $D$ when all the helicity and alignment angles are calculated using the method proposed in Ref.~\cite{Aaij:2015tga}. As shown in Fig.~\ref{fig:alignment}, the alignment is performed properly only for half of the events with $\phi_{\Lambda^* \to p K^-} > 0$ if ordering 1 is taken, and for all the events if ordering 2 is used. This is consistent with the numerical calculations discussed in Section~\ref{sec:single_node}, and demonstrate why ordering 2 is the correct choice.

\section{Another use case: The correct ordering-1 based decay amplitude}
\label{sec:another_use}

In the above discussions in this section, the alignment angle $\theta_p$ is fixed to the angle between the $K^-$ and $\psi$ momenta in the proton rest frame, and the validation is made on the particle orderings. There is another use case where the reference particles are determined, and one can get the correct alignment rotation formula by solving Eq.~\ref{eq:align}. For example, if ordering 1 is used for the $\Lambda_b^0 \to \psi p K^-$ amplitude analysis, the correct alignment rotation is exactly the one in Eq.~\ref{eq:ordering1_align} for events with $\phi_{\Lambda^*\to p K^-} > 0$, as shown in Fig.~\ref{fig:alignment}. For the rest events with $D=8$, Eq.~\ref{eq:ordering1_align} is almost the correct solution, except for an additional minus sign. This can be solved by changing $\theta_p$ to $\theta_p + 2\pi$ for events with $\phi_{\Lambda^*\to p K^-} < 0$. This non-global $2\pi$ factor can recover the non-global minus sign introduced by a improper particle ordering mentioned in Sec.~\ref{sec:single_node}.   

For decays with more than three final-state particles, the alignment rotation obtained from Eq.~\ref{eq:align} can be more complicated, which needs to be expressed using both the polar and azimuthal angles. 
The rotation matrix should be transfered into the Euler style 
under a corresponding Cartesian coordinate system, which makes it easier to get correct  alignment term in the decay amplitude. 
Let's take Fig.~\ref{fig:helicity angle} as an example, and consider the  alignment of particle 1. The Euler angles should be calculated in the $(x_1, y_1, z_1)$ system, and the Euler rotation for the spin-axis alignment can be written as 
 \begin{equation}
R_{z_1}(\alpha) R_{y_1}(\beta) R_{z_1}(\gamma), 
 \end{equation}
The direction of $\vec{x}_1, \vec{y}_1$ and $\vec{z}_1$ can be derived using the four momentum information~\cite{Aaij:2015tga}, based on which the Euler rotation connecting the $(x_1, y_1, z_1)$ and $(x, y, z)$ frames can be obtained, labelled as $R_{\rm Euler, trans}$. Then we have 
\begin{equation}
R_{z_1}(\alpha) R_{y_1}(\beta) R_{z_1}(\gamma) = R_{\rm Euler, trans} R_{z}(\alpha) R_{y}(\beta) R_{z}(\gamma) R^{-1}_{\rm Euler, trans}
\end{equation} 
and 
 \begin{equation}
R_z(\alpha) R_y(\beta) R_z(\gamma) = 
\left(
\begin{array}{cc}
e^{-i(\alpha+\gamma)/2} \cos(\beta/2) & -e^{-i(\alpha-\gamma)/2}  \sin(\beta/2)  \\ e^{i(\alpha-\gamma)/2} \sin(\beta/2) & e^{i(\alpha+\gamma)/2} \cos(\beta/2) \\
\end{array}
\right).  
 \end{equation}
If the alignment rotation obtained from Eq.~\ref{eq:align} is 
\begin{equation}
    R_{\rm Euler} = 
\left(
\begin{array}{cc}
a & b \\ 
c & d  \\
\end{array}
\right),
\end{equation}
we can translate it into 
\begin{equation}
  R^{-1}_{\rm Euler, trans}  R_{\rm Euler} R_{\rm Euler, trans} = 
\left(
\begin{array}{cc}
a' & b' \\ 
c' & d'  \\
\end{array}
\right),
\end{equation}
and the Euler angles for the alignment term can be determined using the following relations  
\begin{equation}
    \begin{split}
        \cos(\beta/2) = |a'|&, \sin(\beta/2) = |b'|, \\
        \alpha+\gamma &= -2{\rm Arg}(a'), \\
        \alpha-\gamma &= 2{\rm Arg}(c'). 
    \end{split}
\end{equation}

\section{Suggestions for the $\Lambda_b^0 \to \psi p K^-$ amplitude analysis}
Given the discussions above, we suggest several modifications on the $\Lambda_b^0 \to \psi p K^-$ decay amplitude with respect to the one used in Ref.~\cite{Aaij:2015tga}, based on either ordering 1 or ordering 2. 

If ordering 2 is used, the modifications are: 
\begin{itemize}
\item For the $\Lambda^*\to p K^-$ decay, use the proton as the reference particle 
\item Add the particle-two factors in the decay amplitude,  including a term of $(-1)^{J_p - \lambda_p}$ to the $P_c$ chain's amplitude, and a term of $(-1)^{J_{\psi} - \lambda_{\psi}}$ to the $\Lambda^*$ chain's amplitude. 
As pointed out in Sec.~\ref{sec:particle-two-factor}, the suggested rotation axis to consider the particle-two factor in this article differs from the one used in Ref.~\cite{Aaij:2015tga}, so the decay angles of the $\psi\to\mu^+\mu^-$ process in the $\Lambda^*$ chain should be modified accordingly, where the  $\phi$ angle should be changed to $\phi+\pi$ when $\phi<0$, and $\phi-\pi$ when $\phi>0$. The modification of the $\phi$  angle definition is just for consistency, but will not  influence the major results of the amplitude analysis, as the additional $\pm \pi$ phase just contributes to a global minus sign to the $\Lambda^*$ chain's amplitude~\cite{Aaij:2015tga}. 
\end{itemize}

If ordering 1 is used, the modifications are: 
\begin{itemize}
\item For events with $\phi_{\Lambda^*\to p K^-} < 0$, change the alignment angle $\theta_p$ to $\theta_p + 2\pi$
\item The operator $R(\pi, \vec{p}_K^{\Lambda^*})$ in Eq.~\ref{eq:ordering1_align}  should also be considered as part of the alignment rotation. The alignment term between the $P_c$ and $\Lambda^*$ chains should be changed from $d_{\lambda_p, \lambda'_p}^{\frac{1}{2}}(\theta_p)$ to $D_{\lambda_p, \lambda'_p}^{\frac{1}{2}}(\pi,\theta_p,0)$. 
\end{itemize}

\section{Conclusion}
The partial-wave amplitude analysis plays an important role for investigating the properties of the resonant structures in multi-body decays. The helicity formalism, as a widely-used technique to construct the decay amplitude, has been adopted to  successfully discover or precisely measure the properties of both exotic and conventional resonant states. However, the principle of choosing the reference decay product when calculating the helicity angles of two-body decays, namely the particle ordering issue, is not often discussed in the traditional usage of the helicity formalism.

In this article, we have first demonstrate the necessity of carefully considering the particle ordering issue, especially for the decays involving spin-half-integral particles where the choice of the reference particles has a non-negligible influence on the interference term. Then a new technique has been proposed to validate whether the decay amplitude is correctly written under a dedicated particle ordering.
This technique checks if the rotation operators involved in different decay chains properly align the spin axes of final-state particles.
A dedicated representation for the operators has been proposed, which could help experimentalists to do event-by-event check of the final-state alignment in a  numerical way. 
Using this new technique, a proper final-state alignment can be reached with any given particle orderings, and 
the inconsistency between different orderings can be  cancelled by assigning different alignment rotation operators. 
Numerical calculations using the  simulated $\Lb \to \psi p \Km$ decays~\cite{Aaij:2015tga} have also been shown as an example. 
The technique proposed in this article will help the ongoing and future particle-wave amplitude analysis of decays with baryons, for example $\Lambda_c^+ \to p K^- \pi^+$, $\Lambda_b^0 \to D^0 p \pi^-$, $B^0\to D^0 p \bar{p}$ and $\psi(2S) \to \eta p \bar{p}$, to construct the decay amplitude in a correct way. 

\clearpage

\section*{Acknowledgements}
We thank Mikhail Mikhasenko and Tomasz Skwarnicki for the valuable discussions about the Dalitz-plot decomposition formalism, which inspires the observation of the particle ordering issue and motivate us to further investigate the helicity formalism, and also for their highlighting to the Jacob-Wick particle-2 phase convention in the amplitude analysis. We thank Andy Beiter, Chen Chen, Mikhail Mikhasenko, Alessandro Pilloni, César Fernández Ramírez, Adam Szczepaniak, Tomasz Skwarnicki, Zhihong Shen and Zehua Xu for the nice discussions about the helicity formalism.

\bibliographystyle{LHCb}
\bibliography{main.bib}

\end{document}